\documentclass[12pt]{spieman}  
\usepackage{amsmath,amsfonts,amssymb}
\usepackage{graphicx}
\usepackage{setspace}
\usepackage{tocloft}
\usepackage{lineno}
\usepackage{soul}

\newcommand{\GC}[1]{\left[#1\right]}
\newcommand{\GA}[1]{\left\{#1\right\}}

\newcommand{\NA}{\mathcal{N}_a}

\newcommand{\NO}{\mathcal{N}_o} 
\newcommand{\NPH}{\mathcal{N}_{ph}} 

\newcommand{\bfh}{\boldsymbol{h}}
\newcommand{\bfi}{\boldsymbol{i}}

\newcommand{\bfalpha}{\text{\boldmath\ensuremath{\alpha}\unboldmath}}
\newcommand{\bfbeta}{\text{\boldmath\ensuremath{\beta}\unboldmath}}

\newcommand{\rev}[1]{} 

\newcommand{\mapolicebackref}[1]{
    \hspace*{\fill} \mbox{\textit {\small #1}}
}
\newcommand*{\backref}[1]{}
\newcommand*{\backrefalt}[4]{%
\ifcase #1 \mapolicebackref{pas de citations}
    \or \mapolicebackref{Cité page #2}
    \else \mapolicebackref{#1 citations pages #2}
\fi
}

\title{Cophasing multiple aperture telescopes with Linearized Analytic Phase Diversity (LAPD)}

\author[a,*]{S. Vievard}\author[a]{A. Bonnefois}\author[a]{F. Cassaing}
\author[a]{J. Montri}
\author[a]{L. M. Mugnier}
\affil[a]{ONERA/DOTA, Universit\'e Paris Saclay, F-92322 Ch\^atillon - France}
\affil[*]{Now in National Astronomical Observatory of Japan, Subaru Telescope, 650 N. A’ohoku Pl, Hilo, HI, 96720, U.S.A.}

\affil[*]{Corresponding author: vievard@naoj.org}

\cftpagenumbersoff{figure}
\cftpagenumbersoff{table} 
\begin{document} 
\maketitle

\begin{abstract}
Focal plane wavefront sensing is an appealing technique to cophase multiple aperture telescopes. Phase
diversity, operable with any aperture configuration or source extension, generally suffers from high computing
load. In this Letter, we introduce, characterize and experimentally validate the LAPD algorithm, based on a
fast linearized phase diversity algorithm \rev{with a capture range comparable to classic phase diversity.} We
demonstrate that a typical performance of $\lambda/75$~RMS wavefront error can be reached.
\end{abstract}


\begin{spacing}{1}   

\section{Introduction}\label{LAPD-intro}

Segmentation of the primary mirror is an inevitable step for space or ground-based optical telescopes in order to maximize the pupil diameter and therefore increase their spatial resolution and collecting power. This was implemented for the Keck telescopes~\cite{matthews-1stDiffractionKECKimages-pasp96} on the ground, or the soon-to-be-launched James Webb Space Telescope (JWST)~\cite{gardner-jwst-ssr06}, and it prevails in the design of the future extremely large optical telescope projects~\cite{Sanders2013,gilmozzi2007european}. For other projects~\cite{villalba2020review}, the goal is to minimize the volume at launch.



A critical sub-system of such instruments is the Cophasing Sensor (CS), whose goal is to measure low order aberrations (i.e. local piston tip and tilt) corresponding to relative positioning errors of the sub-apertures. We make the assumption that the optical quality of each sub-aperture is sufficient so that the high order aberrations are negligible; for example, high technology readiness level of requirements for the JWST segments is a mirror quality with less than 23.7nm RMS error~\cite{10.1117/12.730852}. Several steps are usually required to bring the instrument from a potentially very disturbed state (where images of each sub-aperture are scattered in the focal plane) to the optimal phased state with phasing errors within a small fraction of the operation wavelength (typically below $\lambda/40$~RMS according to a typical error budget for multi-aperture systems~\cite{Harvey90}). Therefore, different algorithms are being investigated to correct for aberrations during each step~\cite{feinberg2007trl}.

For the fine phasing step, Phase
Diversity (PD)~\cite{Gonsalves-a-82,Mugnier-l-06a} is a very appealing and much
studied CS~\cite{Paxman-88,Redding-p-98,Cassaing-p-03a,Mugnier-p-04,korkiakoski2014fast,meimon2008phasing,Mugnier-p-05b,egron2016james} for several reasons: firstly, this CS can use the scientific
camera; this both simplifies the hardware and avoids differential
aberrations between the scientific sensor and the CS. Secondly, it is
appropriate for an instrument with a large number of sub-apertures, because
the complexity of the hardware does not scale with the number of
sub-apertures and remains essentially independent of it, contrarily to the
case of common pupil-plane sensors~\cite{Cassaing-p-03a}. 
Finally, even if this letter only addresses the case of an unresolved object, it can be used on very extended objects, for which it is actually the only reasonable~CS~\cite{Mugnier-p-05b}. However, since the link between the aberrations and the focal and
extrafocal images that are analyzed is highly non-linear, PD usually requires
time-consuming, high computational cost, iterative minimization algorithms. 

Before PD can be applied, the sub-apertures must be coarsely (geometrically) aligned. For this step, Vievard et al.~\cite{Vievard-a-17} developed the ELASTIC algorithm. The latter allows to reduce large tip/tilt aberrations
over sub-apertures from several $\lambda$~RMS to less than $\lambda/8$~RMS when observing an unresolved target. 
In this case, Moc\oe{}ur et al.~\cite{Mocoeur-a-09b} proved through simulations that usual PD
equations could be linearized, and the algorithm thus made much simpler, and
incomparably quicker for whichever observed object (resolved or unknown and unresolved). Other methods, like Fast and Furious~\cite{korkiakoski2014fast, bos2020sky}, also exploited the small aberration regime to implement a fast estimator for wavefront sensing and control but it does not allow open loop estimation, and is limited to unresolved sources.

In this Letter we present a new algorithm similar to the Newton-Raphson method~\cite{ypma1995historical} and based on Moc\oe{}ur et al.~\cite{Mocoeur-a-09b}. We derive and discuss the algorithm, numerically characterize it, and finally we present the first lab experimental validation of linearized PD on a multi-aperture telescope.





\section{The linearized phase diversity principle}\label{LAPD-principle}

The idea behind the new algorithm is to iteratively apply the linearized PD~\cite{Mocoeur-a-09b} around the last estimated set of aberrations. Concretely, it means that a first set of aberrations would be estimated and then used as starting point to derive a new estimate. A few iteration of this method is expected to help increasing the capture range of the linearized PD. 

Following Moc\oe{}ur et al.~\cite{Mocoeur-a-09b} we begin by performing a $1^{rst}$ order Taylor expansion of the instrument Point Spread Function (PSF) $\bfh$, a vector defined on $N_f$ pixels, around
the last known aberration state $\mathbf{a'}$ as a function of the current aberration vector $\mathbf{a}$. With $\delta_{\textbf{a}} = \textbf{a}-\textbf{a'}$, we can write~:
\begin{equation}
\label{linearisation2}
\bfh(\mathbf{a}) = \bfh(\mathbf{a'})+\operatorname{\mathbf{J}}_{\bfh}(\mathbf{a'})\delta_{\textbf{a}} + o(\delta_{\textbf{a}})
\end{equation}
where the vector of the Zernike coefficients of the residual perturbations we want to measure is $\mathbf{a} = (a_{0,0},a_{0,1},...,a_{k,n})$, with $k\in\GA{1,2,3} $ the Zernike mode and $n$ the sub-aperture index. 
$\operatorname{\mathbf{J}}_{\bfh}(\mathbf{a'})$ is the Jacobian matrix of $\bfh$ in $\mathbf{a'}$: $\operatorname{\mathbf{J}}_{\bfh}=\left (\dfrac{\partial{\bfh}}{\partial{a_{0,0}}}, \dfrac{\partial{\bfh}}{\partial{a_{0,1}}}
  ...,\dfrac{\partial{\bfh}}{\partial{a_{k,n}}} \right )$.\\

Doing so, Moc\oe{}ur showed that the best maximum likelihood estimator is the one that minimizes a criterion $L$, defined from the two diversity images $\bfi_1$ and $\bfi_2$, whose Fourier Transform (FT) are written as $\tilde{\bfi}_1$ and $\tilde{\bfi}_2$, as:
\begin{align}
L(\mathbf{a}) &= \dfrac{1}{2\sigma^2} \sum_{\nu}\label{critere}
          |\mathbf{A}(\nu)\mathbf{a} - \mathbf{B}(\nu)|^2+\mbox{Cst}
\end{align}

where $\sigma$ is the noise variance in each image, supposedly the same in both,  $\nu$ is the spatial frequency index, and
\begin{center}
$\mathbf{A}(\nu)=
			\dfrac{ \tilde{\bfi}_2(\nu)\bfalpha_1(\nu)- 
                     \tilde{\bfi}_1(\nu)\bfalpha_2(\nu)}{\sqrt{|\bfbeta_1(\nu)|^2+|\bfbeta_2(\nu)|^2+\epsilon}}$ \mbox{  and  }
$\mathbf{B}(\nu)=
                        \dfrac{\tilde{\bfi}_1\bfbeta_2(\nu)- 
                         \tilde{\bfi}_2(\nu)\bfbeta_1(\nu) }{\sqrt{|\bfbeta_1(\nu)|^2+|\bfbeta_2(\nu)|^2+\epsilon}}$
\end{center}
with, for $i\in\{1,2\}$ (each diversity plane), $\bfalpha_{i}=FT[\operatorname{\mathbf{J}}_{\bfh_i}] \mbox{  and  } \bfbeta_{i} =FT[\bfh_{i}]$.\\

$\mathbf{A}(\nu)$ is a $(k_{max}\NA)$ vector (with $k_{max}$ the highest Zernike mode considered and $\NA$ the number of sub-apertures) defined for each $\nu$ value. We can concatenate all $\mathbf{A}(\nu)$ vectors to obtain a $\mathbf{A}$ matrix with $(N_f, k_{max}\NA)$ dimensions. $\mathbf{A}$ corresponds to the part of the criteria that explicitly depends on the sought aberrations. $\mathbf{B}(\nu)$ is a scalar and is used to compute the vector (with a length of $N_f$) $\mathbf{B}$ corresponding to the criteria expression when there are no aberrations. $\epsilon$ is set to $10^{-6}$, allowing the criteria not to diverge for high frequencies. 

Since $L$ is quadratic, the derivation of its gradient with respect to the aberrations leads to a simple linear equation depending on $\mathbf{a}$. Therefore, the latter can be estimated analytically thanks to the least square solution for complex data:
\begin{equation}
\label{soluce}
\hat{\mathbf{a}} = \left[ \Re(\mathbf{A}^H\mathbf{A})\right]^\dagger . \left[\Re(\mathbf{A}^H\mathbf{B}) \right]
\end{equation}
with $\Re$ the real part and $\dagger$ the matrix generalized inverse. 

\begin{figure}[!h]
    \centering
	\includegraphics[width=0.5\columnwidth]{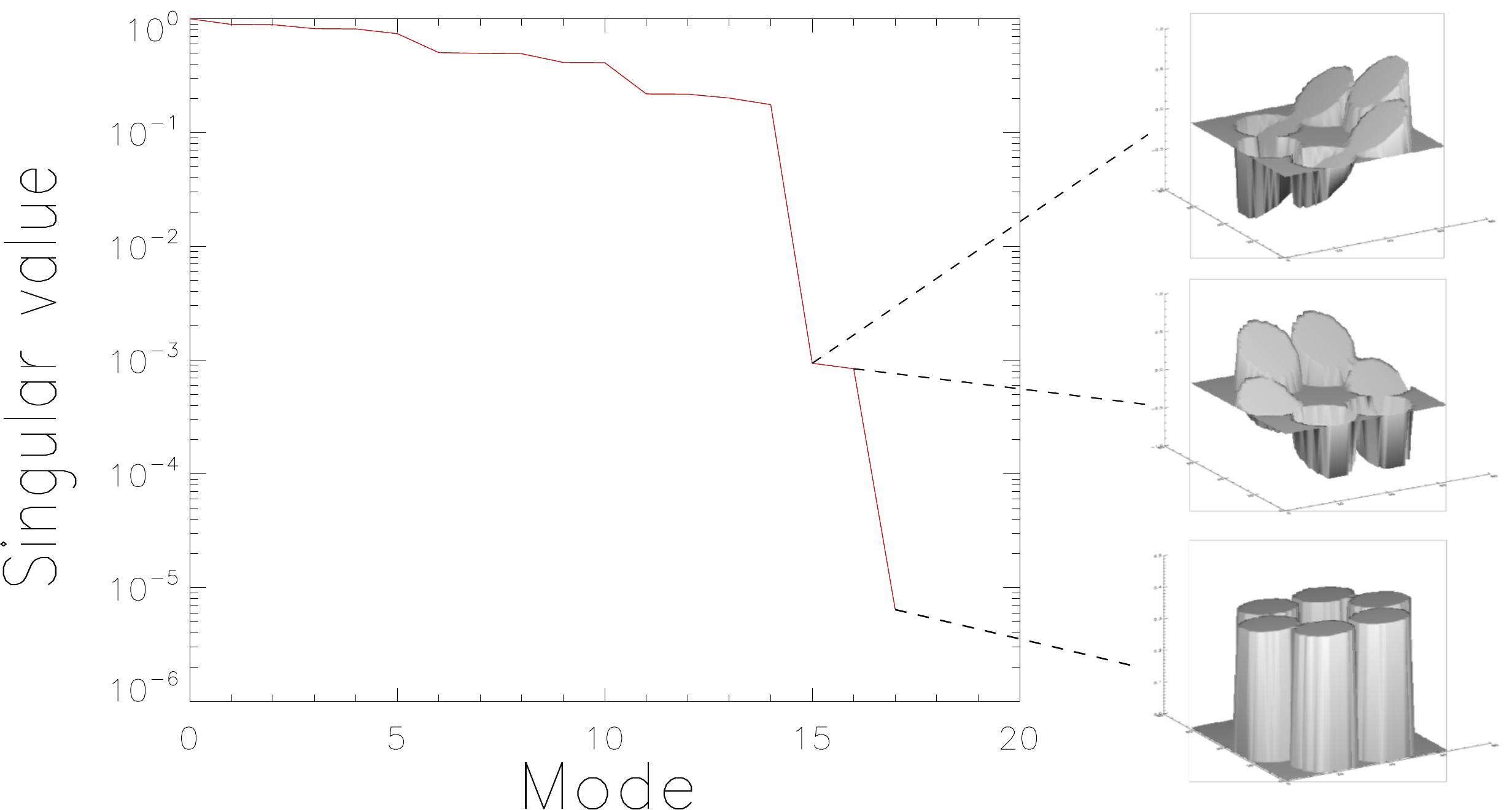}
	\caption{Singular values associated with each modes. The three lowest singular values showed for a 6 circular sub-aperture system correspond to global tip, global tilt and global piston. }\label{lapd_modes}
\end{figure}

Here, we compute the latter with the Singular Value Decomposition. In this process, we can filter out singular modes whose singular values are too small and lead to noise amplification in the inversion. We show Fig.~\ref{lapd_modes} the filtered modes for a 6 circular aperture instrument. We can see that the filtered modes, not seen by the system, are as expected the global piston, tip and tilt.

{}

We found that
range and accuracy could be notably increased by performing a few internal iterations (typically from 1 to 3 as we will show in the following section): a first set of aberrations (noted $\mathbf{a'}$) is estimated, and used as initial guess for a new estimation. It means that the Taylor expansion is then performed around $\mathbf{a'}$ (Eq.~(\ref{linearisation2})). 
We call this algorithm LAPD for Linearized Analytical Phase Diversity.

\section{Numerical validation}\label{sec-num-val}

The optimization and performance evaluations of LAPD are based on a compact pupil with 18~circular sub-apertures placed on a hexagonal grid (see Fig.~\ref{Divopt-LAPD-nirta18} - left) and compared to a "classic" iterative PD algorithm~\cite{Mugnier-l-06a} for an observation of an unresolved source.
All our tests are performed with a simulation tool that simulates the pupil, the phase map error in the pupil plane and that generates images of a point source using the Fraunhofer diffraction theory~\cite{born2013principles} . The size of the images is $64$ by $64$ pixels, with photon noise and a read-out-noise of $5$ photo-electrons. For an initial purpose of metrology tests we will consider high illumination of $5\times10^5$ photo-electrons per image corresponding to a Signal-to-noise ratio of around 700. No higher order aberration are taken into account in our tests here. \rev{All the wavefront errors in this letter are RMS values.}


\subsection{Phase diversity optimization}
\begin{figure}[!h]
    \centering
    \includegraphics[width=0.9\linewidth]{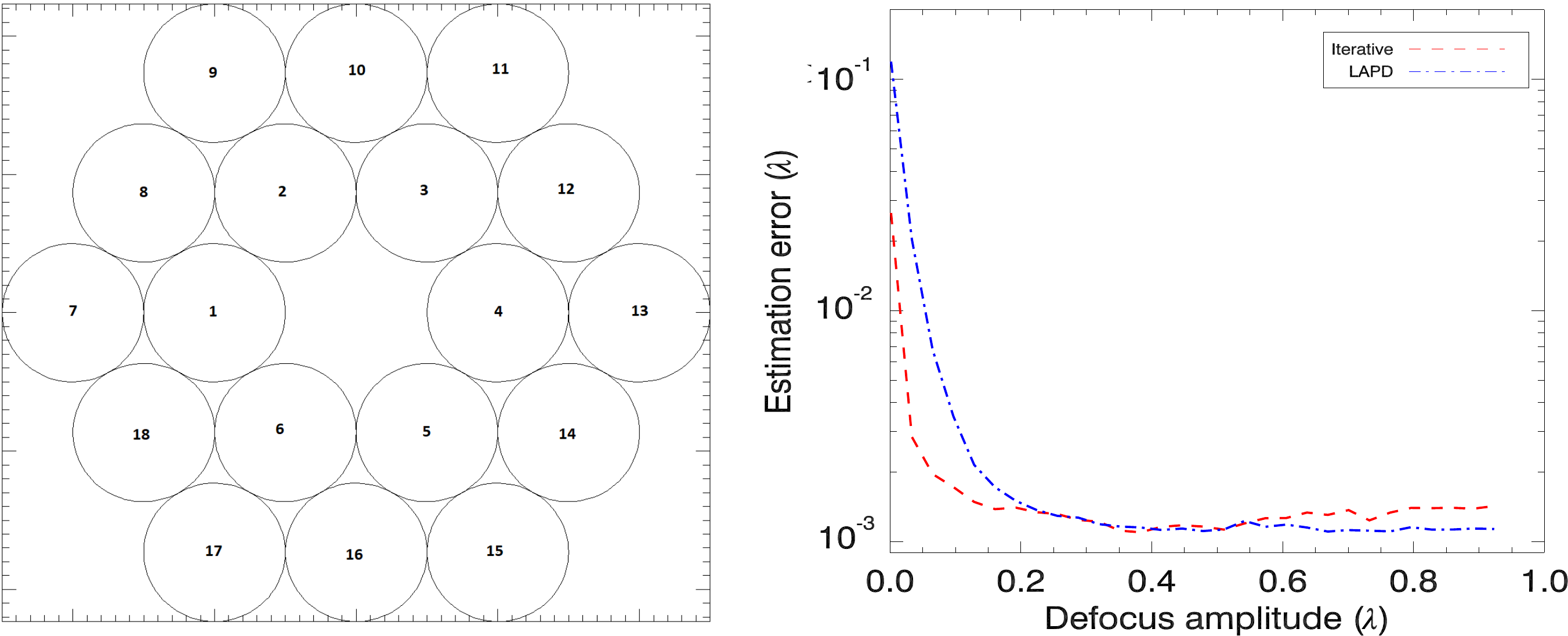}
	\caption{Left: Simulated 18 compact circular sub-apertures. Right : Defocus amplitude optimization for LAPD and the classic PD.} \label{Divopt-LAPD-nirta18}
\end{figure}

We start with the optimization of the defocus introduced between the focal plane image and the diversity image. PD algorithms are based upon the difference between those two images. If this difference is not large enough, performance of such algorithms is poor. In order to find the optimal defocus for LAPD, we fix a random piston/tip/tilt aberration set with a 0.1~rad~RMS standard deviation and study the estimation error for various defocus values. For each value of defocus we perform $\NO=50$ noise and perturbation draws. Fig.~\ref{Divopt-LAPD-nirta18} shows the evolution of LAPD and classic PD algorithm estimation root mean square error (RMSE) in regards of the defocus between images.

We can see that the tendency of both algorithms is that their estimation error decreases when the defocus increases, until a minimal plateau. The latter minimum is slightly above $\lambda/1000$~RMS for both algorithm when the defocus amplitude is larger than $0.2\lambda$. This result is in adequation with previous work~\cite{lee1999cramer} on optimal defocus diversity in the case of a monolithic telescope. In the following we will use a $0.3\lambda$ defocus amplitude.

\subsection{Linearity of the sensor}
\begin{figure}[!h]
\centering
	\includegraphics[width=0.9\linewidth]{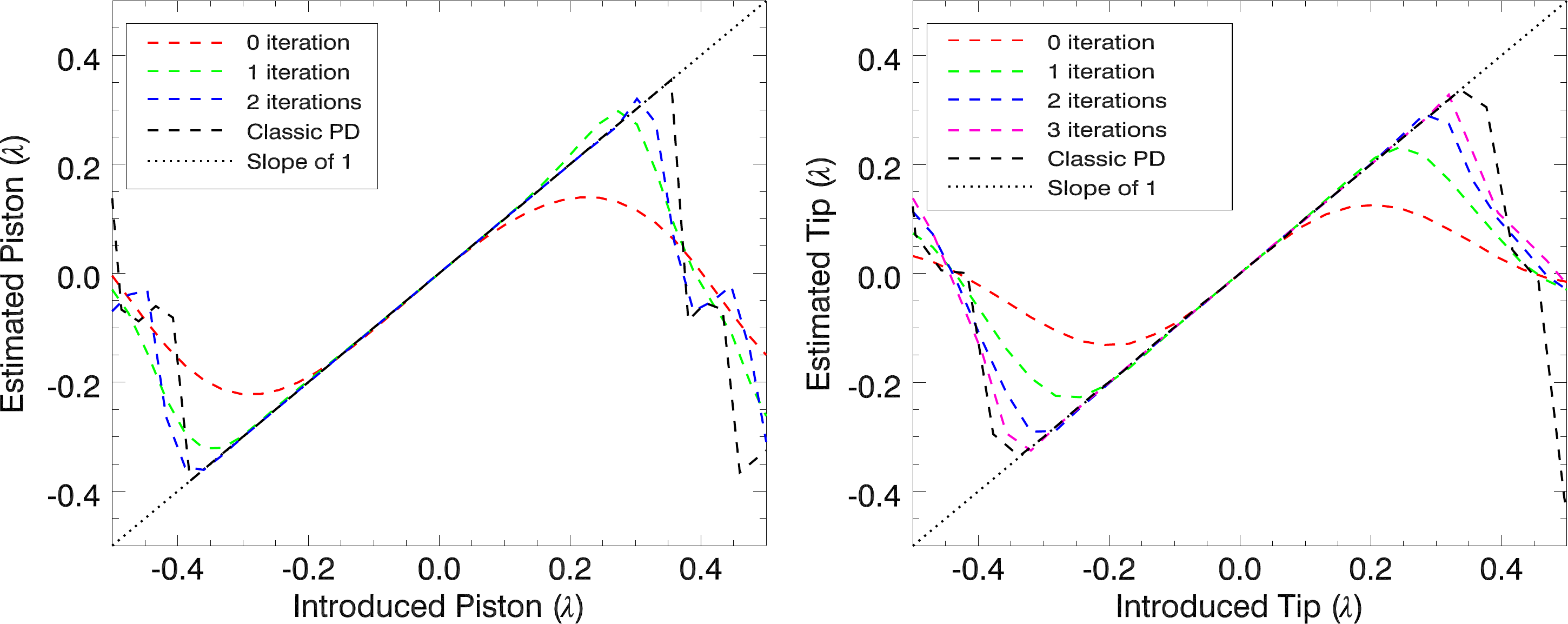}
	\caption{Evolution of the piston (\rev{right}) and tilt (\rev{left}) estimation as a function the introduced piston or tilt.}\label{Linearite-NIRTA18-iter-P}
\end{figure}

Let us now study the linearity and dynamic range of LAPD. 
We study the LAPD estimation using $0$ to several internal iterations for a piston and a tilt error range of $\GC{-0.5\lambda;0.5\lambda}$. Fig.~\ref{Linearite-NIRTA18-iter-P} shows the evolution of the mean estimation over $\NO=50$ noise draws as a function of the introduced piston or tilt. We keep an illumination of $5\times10^5$ photo-electrons per image for this test.

On Fig.~\ref{Linearite-NIRTA18-iter-P} (left) we can see that LAPD without internal iteration has a linear response on the $\GC{-0.2\lambda;0.1\lambda}$ piston range. Increasing the number of internal iterations enlarges LAPD linear range to about $\GC{-0.4\lambda;0.35\lambda}$. More than 3~internal iterations do not increase the dynamic range. We also over-plot the linear range of the classic PD $\GC{-0.45\lambda;0.4\lambda}$. The asymmetry that we notice in the linear range comes from the choice of the small unilateral defocus as a phase diversity. Indeed, further tests showed that the linear range will be symmetric only in the case of a symmetric diversity (two diversity images with opposite defocus) or by increasing the defocus to several $\lambda$s. On Fig.~\ref{Linearite-NIRTA18-iter-P} (right) we see that LAPD tilt estimation is linear on the $\GC{-0.1\lambda;0.1\lambda}$ range without internal iteration. When increasing the number of internal iterations to 3 the tilt estimation is linear on the $\GC{-0.32\lambda;0.32\lambda}$ range, just like the classic PD. We then showed that LAPD piston or tilt estimation is linear on a given dynamic range and the latter can be increased by using more internal LAPD iterations. Doing so, LAPD dynamic range is comparable to classic phase diversity algorithm's for tilt estimation. The dynamic range of the piston estimation does not reach the same as the classic PD, however we can assume that static phasing errors would not be as large during the telescope \textit{fine phasing} step.

\subsection{Noise propagation study}
\begin{figure}[h!]
 \centering
    \includegraphics[width=0.9\linewidth]{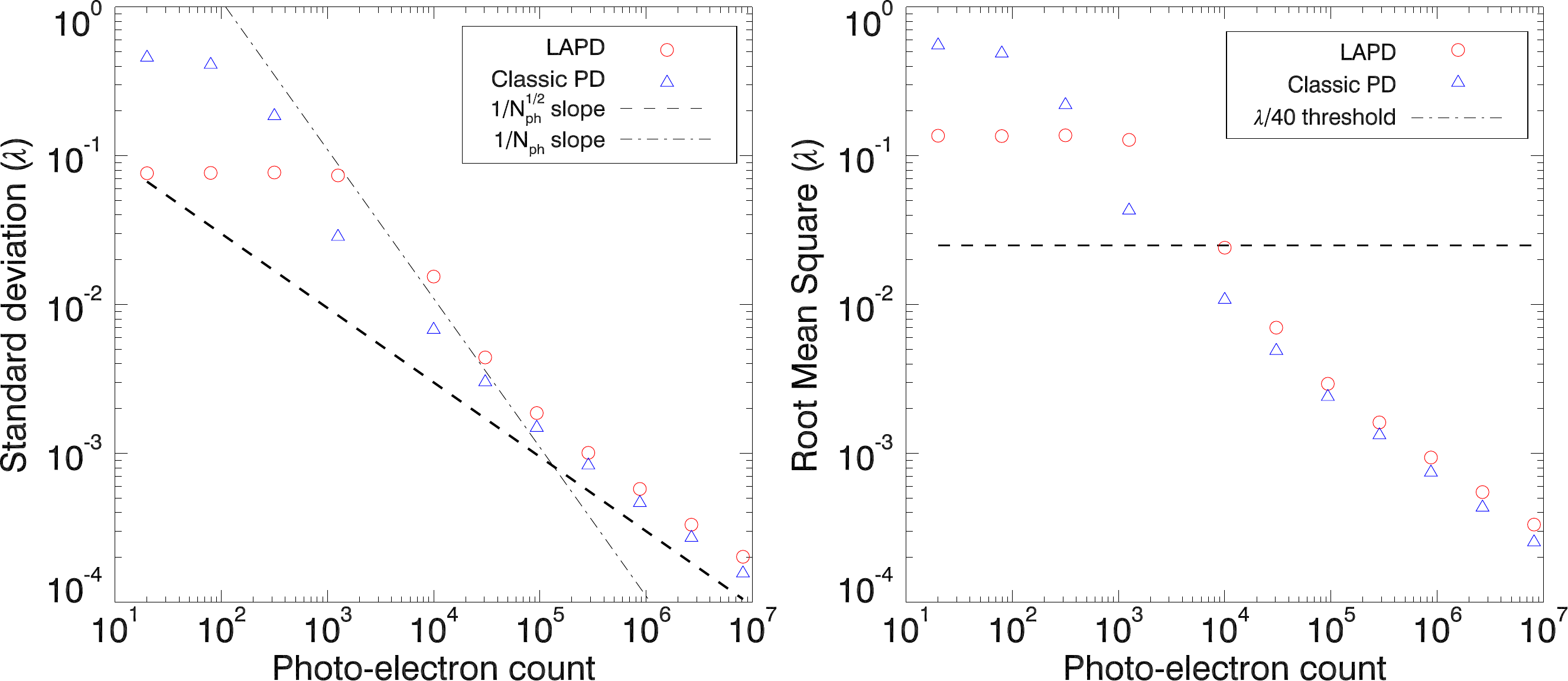}
\caption{LAPD \rev{noise} propagation study. Evolution of the standard deviation (left) and RMSE (right) of the estimation as a function of image illumination.}\label{Propnoise-LAPD-nirta18}
\end{figure}

Finally, we study the performance of LAPD with regards to the photo-electron count ($\NPH$). We set the number of internal iterations to two, as we saw previously that it is sufficient to significantly increase the capture range of the algorithm. For each case we study the standard deviation and the RMSE of the piston/tip/tilt estimation over the $\NO=50$ draws. A random piston/tip/tilt set with a 0.2~rad standard deviation is applied for each estimation. Fig.~\ref{Propnoise-LAPD-nirta18} left and right respectively show the evolution of the standard deviation and the RMSE as a function of the number of photo-electrons in each image.

We can distinguish different regimes in the standard deviation evolution of the LAPD and classic PD estimations: at low flux (below $10^3$~photo-electrons for LAPD, below $5\times10^2$~photo-electrons for the classic PD) the standard deviation of the estimation is constant. The value of the standard deviation is around $\lambda/16$ for LAPD and around $\lambda/5$ for the classic PD. This difference can be explained by the fact that the few number of iterations for LAPD can play the role of regulation~\cite{yao2007early}. Since the classic PD does not have any kind of regularisation, the performance at low flux is significantly worse. At high flux ($\NPH > 10^4$, respectively $\NPH > 10^5$), the slope of detector-noise (respectively shot-noise) regime is clearly evidenced in the case of LAPD estimation standard deviation. In the case of the classic PD, the shot-noise regime is evidenced for $\NPH > 10^4$. 
The evolution of the RMSE informs us that the estimation error is $<\lambda/40$ for an illumination~$>~10^3$~photo-electrons for the classic PD and $>10^4$~photo-electrons in the case of LAPD. Both algorithms have similar behaviors. However we can notice that the classic PD algorithm is slighly more accurate when the flux is above $3\times10^5$. This \rev{sensitivity} difference is reduced (even \rev{overcome}) by using LAPD with more internal iterations\rev{, but leading to a longer computation time}.\\
In this section the LAPD implementation (2~internal iterations, unoptimized IDL code) runs about $3\times$ faster than the classic PD. The main bottleneck is the sequential computation of FTs for $\bfalpha_{i}$. For the future, this should trivially be parallelizable for manycore architectures with little overhead.


\section{Experimental validation}\label{lapd-expval}
In order to experimentally validate LAPD, we use the BRISE bench built by Onera~\cite{cassaing2006brise,Vievard-a-17}. We conduct tests imaging the collimated output of a fibered single-mode fiber near 635~nm, acting like an unresolved source, through six circular apertures of a segmented mirror. Each segment is supported by three piezoelectric actuators, allowing to introduce piston, tip or tilt perturbations. Downstream, a phase diversity module is used to simultaneously form a focused and a defocused image of the object on a $1300\times~1000$ pixels camera, from which we extract two $128\times~128$ diversity images. We test LAPD with a $0.3\lambda$ defocus between the two images, and no internal iteration. 

A closed-loop sequence is performed to correct piston/tip/tilt perturbations since it is not possible to perform a single-step correction, due to not perfectly deterministic active mounts. 
\begin{figure}[!h]
\centering
	\includegraphics[width=0.9\linewidth]{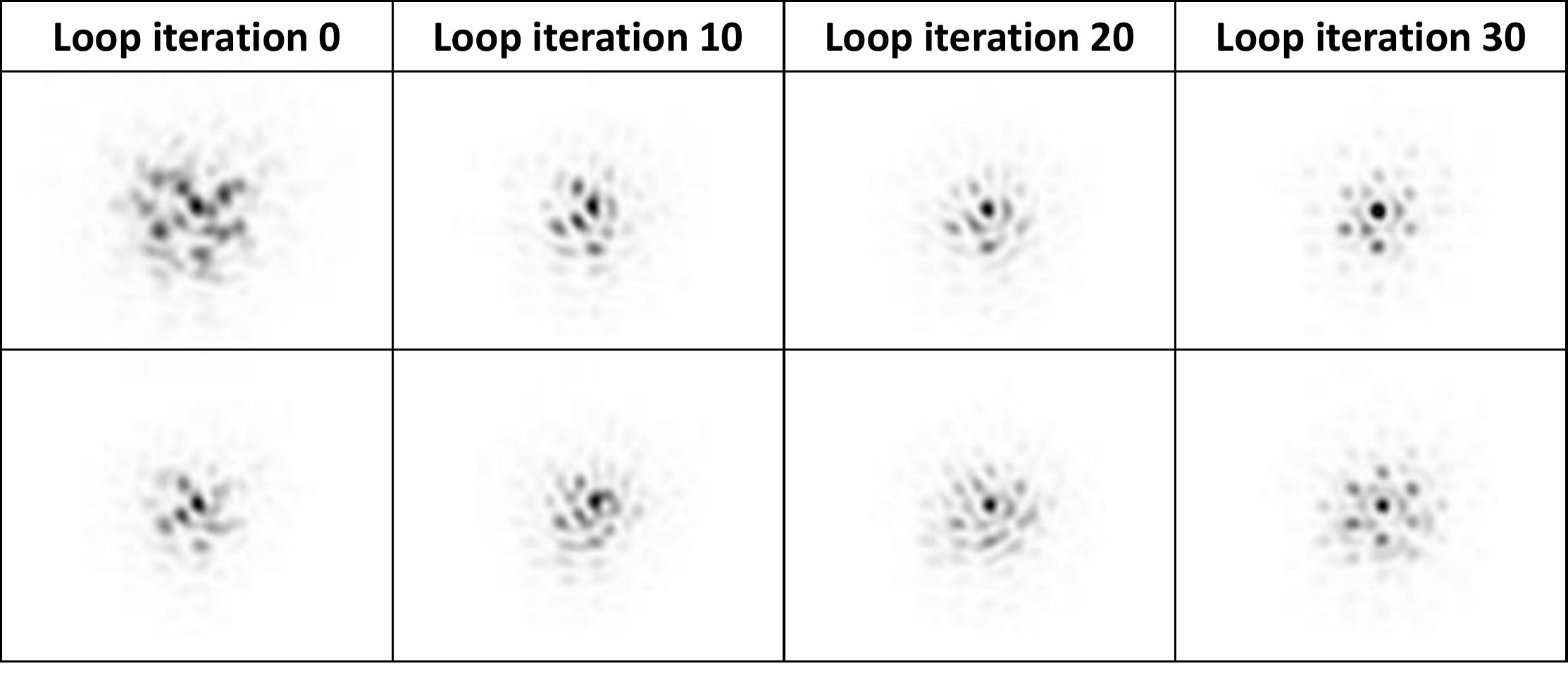}
\caption{Experimental validation of LAPD: focal (top) and defocused (bottom) PSF evolution during iterations of a closed-loop correction on a 6-subaperture mirror bringing the global phase error down to a $\lambda/75$ RMS.}
\label{closeloop-LAPD}
\end{figure}
Figure~\ref{closeloop-LAPD} shows the effect of LAPD correction. The departure point (iteration 0) of the cophasing sequence results from the previous coarse alignment~\cite{Vievard-a-17, 10.1117/12.2312580}, where we can see small piston/tip/tilt errors from the PSF shape. The loop is closed with a pure integrator with $0.2$ gain. We can see over the loop iterations that LAPD manages to estimate the small errors to reach the phased instrument PSF. The estimated piston/tip/tilt errors over the sub-aperture have a maximal amplitude of $0.1\lambda$ at the beginning of the loop closure, and are brought to less than $0.02\lambda$, with a $\lambda/75$ RMS dispersion. The total flux is $10^6$~photo-electrons. A comparison of the obtained dispersion ($\lambda/75$) with the simulation in Fig.~\ref{Propnoise-LAPD-nirta18}-left ($10^{-3}\lambda$) shows a difference of a factor 10. This difference can have several origins mainly linked to the simplified case used for simulation: 1- the error induced by the piezoelectric actuators ; 2- high order aberrations in the optical setup ; 3- bench modelization errors. Even with the hypothetical presence of these errors, LAPD was still able to bring the instrument in a finely cophased state.

\section{Conclusion}
We presented here LAPD, an analytical fast solution to the fine phasing of multi-aperture systems from a pair of diversity images near the focal plane. We showed with simulations on a 18-subaperture imager observing (for simplicity) an unresolved source that LAPD could increase its capture range with a small amount of internal iterations (typically less than 3), and that it can estimate piston/tip/tilt aberrations with an error below $\lambda/40$~RMS for an illumination $>10^4$ for $64\times64$~pixel images. Finally, we validated LAPD on a 6-subaperture mirror. We showed that after correction of large alignment errors with ELASTIC we were able to bring the piston/tip/tilt errors down to less than $0.02\lambda$~RMS using a close-loop enslavement using LAPD. This work could be useful as an easy-to-implement, computationally efficient solution for the fine phasing of any multi-aperture telescope, and could also be used in the case of unresolved objects.

\section*{Acknowledgements}
This research was partly funded by ONERA's internal research projects VASCO and METOPE ; Thales Alenia Space co-funded S.~Vievard PhD thesis; the BRISE bench was funded by French DGA. The authors would like to thank N. Treps for the direction of S.~Vievard PhD; J.-P.~Amans from GEPI of Observatoire de Paris, for the design and
manufacturing of the segmented mirror.


\bibliography{article}   
\bibliographystyle{spiejour}   


\end{spacing}
\end{document}